\documentclass{article}

\usepackage{arxiv}
\usepackage[font={small,it}]{caption}
\usepackage[utf8]{inputenc} 
\usepackage[T1]{fontenc}    
\usepackage{hyperref}       
\usepackage{url}            
\usepackage{booktabs}       
\usepackage{amsfonts}       
\usepackage{nicefrac}       
\usepackage{microtype}      
\usepackage{lipsum}
\usepackage{graphicx}
\usepackage{amsmath}
\usepackage[dvipsnames]{xcolor}
\graphicspath{ {./figures/} }
\title{Physical Accuracy of Deep Neural Networks for 2D and 3D Multi-Mineral Segmentation of Rock micro-CT Images}

\author{
  Ying Da~Wang \\
  School of Minerals and Energy Resources Engineering\\
  University of New South Wales\\
  Sydney, NSW, 2052 \\
  \texttt{yingda.wang@unsw.edu.au} \\
  \And
  Mehdi ~Shabaninejad \\
  Department of Applied Mathematics \\
  Australian National University\\
  Canberra, ACT, 2601 \\
  \texttt{mehdi.shabaninejad@anu.edu.au} \\
  \And
  Ryan T.~Armstrong \\
  School of Minerals and Energy Resources Engineering\\
  University of New South Wales\\
  Sydney, NSW, 2052 \\
  \texttt{ryan.armstrong@unsw.edu.au} \\
  \And
  Peyman~Mostaghimi \\
  School of Minerals and Energy Resources Engineering\\
  University of New South Wales\\
  Sydney, NSW, 2052 \\
  \texttt{peyman@unsw.edu.au} \\
}

\begin{document}
\maketitle

\begin{abstract}
Segmentation of 3D micro-Computed Tomographic ($\mu$CT) images of rock samples is essential for further Digital Rock Physics (DRP) analysis, however, conventional methods such as thresholding, watershed segmentation, and converging active contours are susceptible to user-bias. Deep Convolutional Neural Networks (CNNs) have produced accurate pixelwise semantic segmentation results with natural images and $\mu$CT rock images, however, physical accuracy is not well documented. The performance of 4 CNN architectures is tested for 2D and 3D cases in 10 configurations. Manually segmented $\mu$CT images of Mt. Simon Sandstone are treated as ground truth and used as training and validation data, with a high voxelwise accuracy (over 99\%) achieved. Downstream analysis is then used to validate physical accuracy. The topology of each segmented phase is calculated, and the absolute permeability and multiphase flow is modelled with direct simulation in single and mixed wetting cases. These physical measures of connectivity, and flow characteristics show high variance and uncertainty, with models that achieve 95\%+ in voxelwise accuracy possessing permeabilities and connectivities orders of magnitude off. A new network architecture is also introduced as a hybrid fusion of U-net and ResNet, combining short and long skip connections in a Network-in-Network configuration. The 3D implementation outperforms all other tested models in voxelwise and physical accuracy measures. The network architecture and the volume fraction in the dataset (and associated weighting), are factors that not only influence the accuracy trade-off in the voxelwise case, but is especially important in training a physically accurate model for segmentation.

\end{abstract}


\section{Introduction}
\label{sec:intro}
Digital images of rock samples, such as those obtained from 3D X-ray micro-Computed Tomography ($\mu$C), are commonly used for numerical analysis of the physical properties of the samples \cite{BLUNT2013197,MOSTAGHIMI2017143}. Features captured at the micrometre scale allow the pore structure of reservoir rocks to be accurately characterised \cite{Flannery, Coenen} in a non-invasive and non-destructive manner, so samples may later used experimentally or otherwise \cite{lindquist, Hazlett, Wildenschild, RN7}. After processing and segmentation \cite{iassonov2009segmentation}, such segmented digital rocks are used as is for compositional characterisation by Mineral Liberation Analysis \cite{mla}, and for the analysis of fluid transport properties by numerical simulation of pore-scale transport mechanics for example single-phase flow, particle tracking, reactive transport and multiphase flow \cite{schluter2014image,MCCLURE20141865,wang2019multi,pfvs,DGDD,peymanK,agglomTrai, Krakowska, ferrand1992effect,Mostaghimi2016,LIU2017121}.

Image segmentation techniques used in Digital Rock Physics (DRP) range from simple thresholding \cite{Sezgin2004SurveyOI} to multi-step marker based methods \cite{iassonov2009segmentation,ANDRA201325}. All such methods require a degree of user judgement and subsequent bias in the manual tuning of each step. It is common for the segmentation of a digital rock to consider only the pore space and the solid grain space as the distinguishable phases, treating all minerals within the rock sample as a singular entity with which uniform physical and chemical properties may be modelled. While acceptable for the simple case of single-phase unreactive flow, this simplification is not applicable to all cases, especially in cases such as P and S wave velocity modelling \cite{ANDRA201325}, mixed wetting flow simulation \cite{takaiMultiWet}, and reactive transport \cite{LIU2018130}.

Convolutional Neural Networks (CNNs) built within a Deep Learning framework \cite{SCHMIDHUBER201585} have been used in a variety of image based problems, including image-to-image translation \cite{pix2pix2016,cyclegan}, super resolution \cite{EDSR,SRGANledig}, image classification \cite{resnet}, and semantic segmentation \cite{maskrcnn,ronneberger2015unet}. These techniques manifest themselves in the work flow of many applications, such as face/object detection \cite{facedetection}, autonomous driving \cite{segnet}, speech recognition \cite{speech}, and medical image processing/analysis \cite{ldct1,ldct2,ldct3,umeharaSRCNNmedical,ctmedicalresolution1,circleGAN,segITO,segMed3DLI202075}. Specific to Digital Rock Analysis, CNNs have been used for super resolution \cite{wang2019super,wangedsrgan,DRSRD1,DeepRock-SR,peymanTSR}, and binary and multi-mineral segmentation \cite{computers8040072,KARIMPOULI2019142}. More general approaches using machine learning has also found success in the prediction of petrophysical properties from $\mu$CT images \cite{RN112,RN116,RN115,RN114,RN113} including permeability, porosity, surface area, and other morphological parameters. 

For the specific task of segmentation of $\mu$CT images, multiple challenges and uncertainties exist. The limited availability of registered segmented images and the expense in obtaining a large and broad dataset spanning different rock types is a factor that influences the manner in which segmentation by CNNs is performed. Data augmentation of sparsely available multi-mineral segmentation examples can be done \cite{KARIMPOULI2019142}, or widely available binary segmentation of multiple different rock samples can be used \cite{computers8040072}. In this study, availability of multi-mineral segmented data is not a major issue, as the dataset outlined in Section \ref{sec:Datasets} contains 2.662 Billion voxels, which is comparable to the augmented dataset used by \cite{KARIMPOULI2019142} that contains 1.3 Billion voxels. In photographic cases, examples of datasets for semantic segmentation and object recognition frequently surpass the Billion pixel mark, \cite{imagenet,audi,coco,pascal}. In the training, testing, and validation of neural network performance, the main advantage of a broader and larger dataset is the extrapolative accuracy and reliability of the trained network. In the case of having access to training and testing on multiple samples \cite{computers8040072}, validation sensitivity can be performed, which in this case showed up to 10\% difference in binary segmentation accuracy depending on the training and testing data split. The neural network architectures tested in segmentation of digital rock images include the SegNet \cite{segnet} and the U-Net \cite{ronneberger2015unet} architectures. Of these, SegNet in its original and deeper forms was tested for 2D multi-mineral segmentation, while U-Net was used for binary segmentation in 2D and 3D. One of the most important considerations in image segmentation for the purpose of physics-based analysis (such as in DRP), is the physical accuracy of the segmentation during validation. This measurement of physical accuracy is typically performed by comparing permeability with some reference point (typically experimental) \cite{Leu2014}. While segmentation accuracy in 2D multi-mineral and 3D binary tasks shows high accuracy (over 95\%) as measured using pixel counting, there are no such guarantees in the case of permeability, which tends to be highly sensitive to the segmentation result \cite{yufuseg}. In the case of CNNs which operate as a black-box, ensuring the physical accuracy of the CNN result is a top priority.

An important caveat to address is that the definition of segmentation accuracy is itself subjective. Using a dataset of manually segmented images will result in an inherent bias towards the segmentation as performed by the person generating the data. This issue of segmentation user-bias is an as of yet unaddressed issue in DRP, and is ultimately not within the scope of this study. One simple method of addressing user bias in Multi-Mineral Segmentation is to simply train any one of the neural networks tested in this study on a dataset large and varied enough to average out any biases in segmentation. Regardless, treating the segmented dataset as the ground truth is sufficient to allow a controlled comparison of neural network accuracy.

In this study, multiple neural network architectures are trained and tested for the task of multi-mineral segmentation. A total of 4 architectures are tested in 10 configurations in 2D and 3D, including the SegNet \cite{segnet}, U-Net \cite{umeharaSRCNNmedical}, ResNet \cite{resnet}, and U-ResNet, which is a custom formulation designed as part of this study, similar in its formulation to the MultiResUNet \cite{MultiResUNet}. The resulting testing performance is analysed for all segmented phases. Pixelwise accuracy results show that more advanced network designs that build upon the SegNet architecture outperforms SegNet by over 5\% in some cases, and visual inspection reveals much sharper accuracy at the edges of segmented phases. Physical accuracy is measured by (1) analysis of the phase connectivity using the Euler Number, (2) comparison of permeability calculated from direct flow simulation in the pore space, and (3) comparison of the results from direct simulation of two-phase flow in single and mixed wetting cases. Results show a much larger and random variation in physical accuracy, with the U-ResNet 3D model showing the most consistent and accurate results. The choice of CNN model and the resulting performance of said model should be carefully validated using physical measures in cases where the physics are the focus of further analysis, which is the case in DRP.

\section{Materials and Methods}
\label{sec:Materials and Methods}

\subsection{Datasets}
\label{sec:Datasets}
To test the accuracy of various Neural Network Architectures for the purpose of Multi-Mineral Segmentation, a training and testing dataset comprised of 2D and 3D Mt. Simon sandstone images is used. The data originates from the registration of segmented slices obtained from Quantitative Evaluation of Minerals by SCANning electron microscopy (QEMSCAN) with Energy-Dispersive X-ray Spectroscopy (EDS) \cite{QEMSCAN, LIU201712,qemscanorig}. A miniplug from Mt. Simon reservoir sandstones is cored to approximately 3 mm in diameter and 5 mm in length and $\mu$CT scanned at x-ray facilities in the University of New South Wales. Sample preparation begins with placing the mini-plugs in a vacuum chamber to be evacuated for 20 minutes, followed by vacuum infiltration of a resin mixture for 1 hour. The sample is then pressurized to 250 kPa for one day to dissolute any air bubbles and to allow resin to invade smaller pores. The embedded mini-plug is then removed from its mold and cut crosswise in the middle to create two surfaces. Each cut surface is polished using a Struers Tegramin polishing machine with a series of polishing steps of varying disc roughness and force. This polished surface is then carbon coated to minimize beam charging and thermal damage during Scanning Electron Microscopy (SEM) imaging \cite{shabaninejad2017pore} and then loaded into a Quanta 650 FEG (FEI) microscope to acquire the mineral map and back-scattered electrons (BSEM) of the full cross-section. The scanning is performed at 15 kV from 13 mm using QEMSCAN iMeasure software to acquire a SEM-EDS mosaic image of the full cross-section of the mini-plug at a resolution of 1.8 micrometers. Further image processing such as stiching, noise removal and mineral classification is subsequently performed using QEMSCAN NanoMin software. The 2D BSEM image of each miniplug is then registered into its 3D $\mu$CT tomogram. This then allows the 2D SEM-EDS mineral map to be registered to the corresponding cross- section of the mini-plug \cite{Latham2008ImageRE,LIU201712}. 

\begin{figure}
  \centering
    \includegraphics[width=\textwidth]{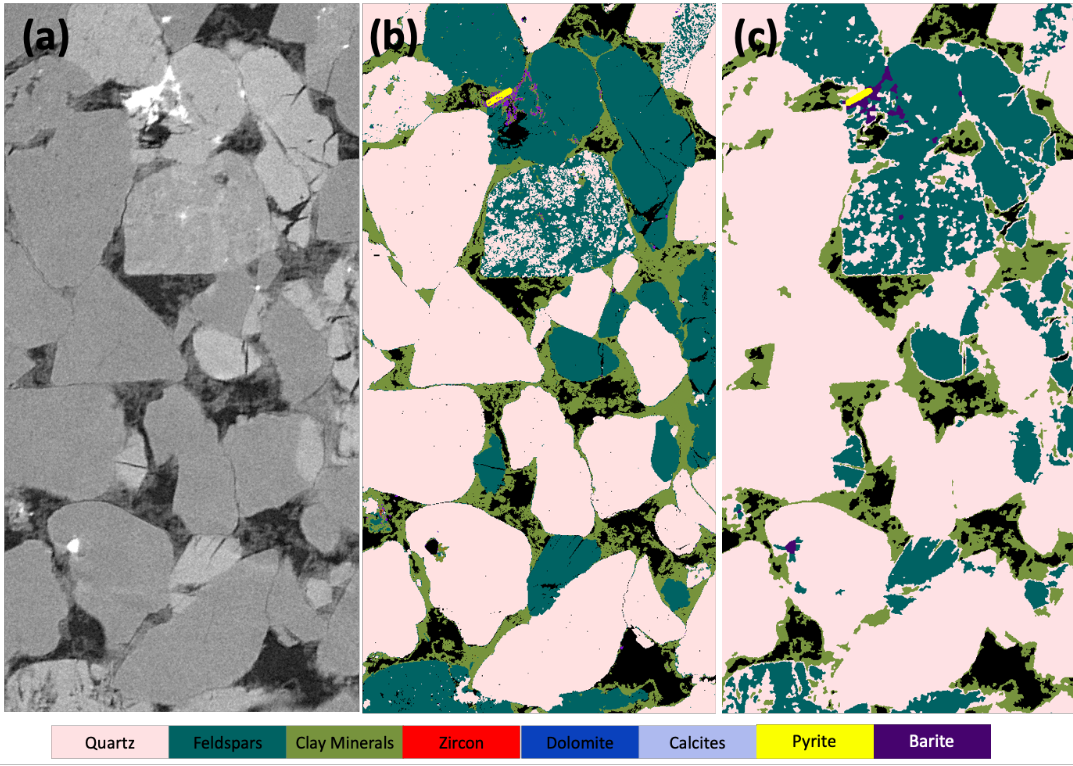}
    \caption{Subarea of a slice of the Mt Simon sample. a) dry micro CT tomogram b) registered 2D mineral map image acquired from QEMSCAN c) mineral segmentation using x-ray intensities of minerals. Size of image is 1800 $\mu$m x 900 $\mu$m.}
    \label{fig:qemscanToCACSeg}
\end{figure}

Following this process, 3D mineral segmentation of the $\mu$CT tomogram is performed by distinguishing the X-ray intensity range of each mineral as guided by the 2D QEMSCAN images registered to the 3D tomogram. Figure \ref{fig:qemscanToCACSeg} shows a subarea of mineral segmentation along with the 2D QEMSCAN and registered dry tomogram. There are some discrepancies between QEM images and mineral segmentation which stem from: 

\begin{itemize}

\item Registration is not always perfect, so small movements within the range of 1-2 voxels impacts the analysis. 
\item Clays (mainly kaolinite) will deform during the cutting and polishing steps, thus pixelwise comparison of deformed clay aggregates will be associated with some errors. 
\item Some minerals can be detached from the polished surfaces during cutting and polishing steps, which can create a false pore space within mineral map image. 
\item Segmentation always comes with some errors due to tomogram quality (mainly phase contrast in high density grains) and the user-bias in choosing the intensity thresholds. 
\item The algorithm used, converging active contours \cite{sheppard2004techniques}, mainly utilises the X-ray intensity of minerals, so it is almost impossible to differentiate between minerals with similar x-ray attenuations (such as Quartz and Na-plagioclase). 

\end{itemize}

While the QEMSCAN data is a direct mapping of identified minerals with their $\mu$CT counterparts, the 2D nature of QEMSCAN data restricts its usage to 2D CNNs only. While the image resolution of QEMSCANs is also quite high (in the sub-micrometre range), thus resulting in a reasonably large number of pixels in the order of 5000px squared, this is 50 times less than the amount of data used in previous studies for training CNNs for digital rock segmentation \cite{KARIMPOULI2019142}. Furthermore, the presence of a representative distribution of image features is not guaranteed in limited planar slices which comprise of less than 100 grains. Thus this method of obtaining 3D multi-mineral segmentation allows the training and benchmarking of both 2D and 3D CNNs on an appropriately sized dataset.

\begin{figure}
  \centering
    \includegraphics[width=\textwidth]{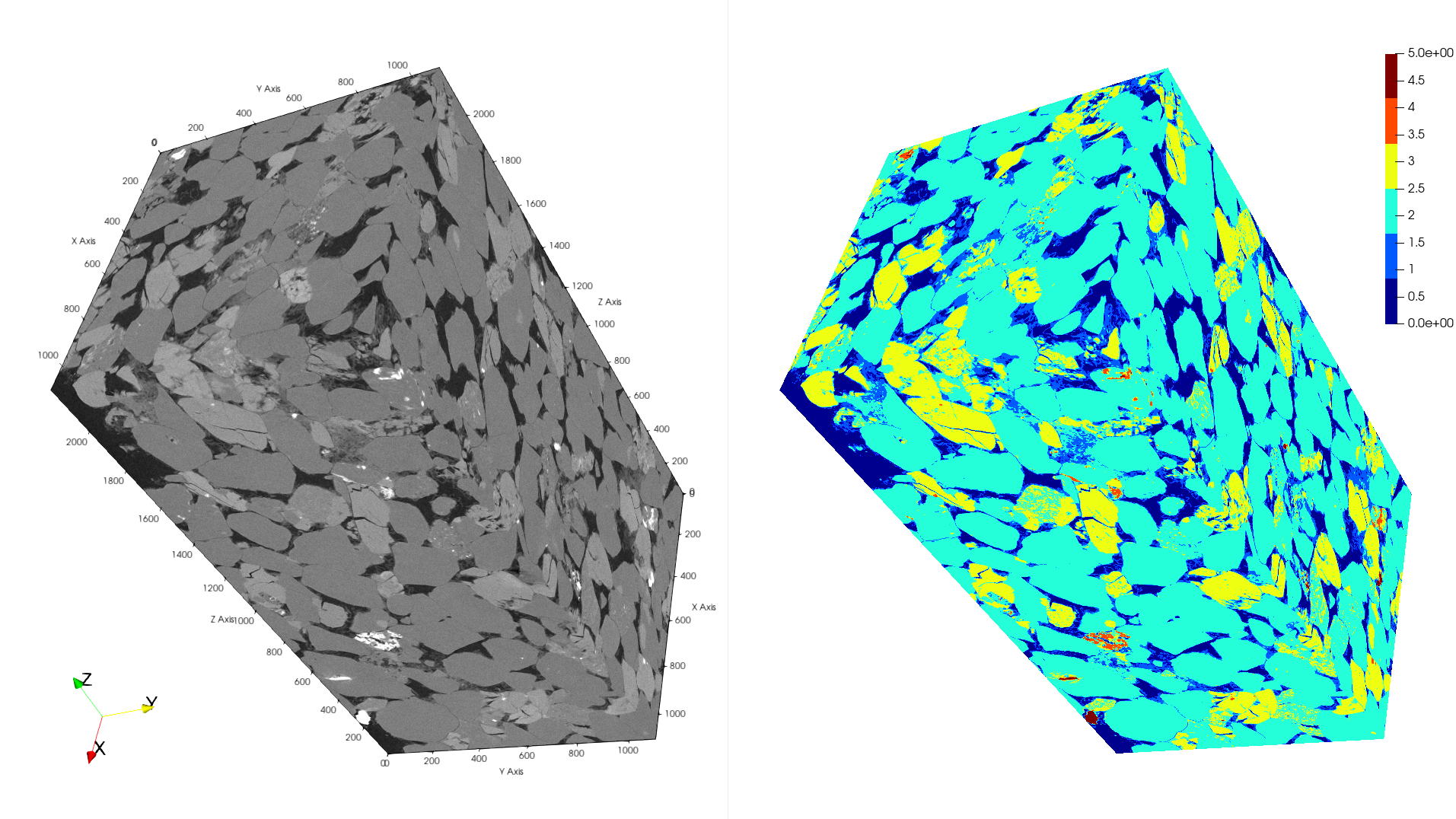}
    \caption{Left: Full size greyscale $\mu$CT image of Mt. Simon rock, measuring 1100x1100x2200 voxels. Right: Sample segmented by registration to QEMSCAN slices, totalling 5 mineral phases + pore space}
    \label{fig:simonFull3DQEM}
\end{figure}

The resulting segmentation is shown in Figure \ref{fig:simonFull3DQEM}, and is comprised of 6 identified phases. In order from label 0 to 5, they are Pore (0), Clay (1), Quartz (2), Feldspar (3), Micas (4), and a mixed group including Sulfides (Pyrite) and other high density minerals (5).

In total, the 2D dataset is comprised of 8,800 greyscale and segmented images, each measuring 550x550 pixels and split into 8,000 for training and 800 for testing. The 3D dataset is comprised of 1,000 volumes (split 80/20 for training/testing), each measuring 128 voxels cubed. 

\subsection{Segmentation Networks and Training Schedules}
\label{sec:segNets}
Network architectures used in this study are loosely based on the encoder-decoder structure, and build upon the original SegNet \cite{segnet} with concepts from ResNet \cite{resnet} and U-Net \cite{ronneberger2015unet}. The networks are constructed in a modular manner, with each increase in model complexity based on the structure of the previous. As is the case in the most effective neural networks, skip connections introduced in ResNet \cite{resnet} are used liberally to improve the preservation of shallow details. The basic structure takes the form of an Encoder-Decoder network based on the SegNet architecture, and from this, ResNet, U-Net, and U-ResNet architectures are created. The details of each network are outlined in their respective sections ahead, and the source code is available at \url{https://github.com/yingDaWang-UNSW/SegNets-3D}. 

Each network is trained an initial learning rate of 1e-4 using the Adam optimiser \cite{AdamKingma}. The batch size used in 2D training is 16x256x256, while the 3D batches are sized as 2x128x128x128. Training is terminated once validation accuracy reaches a plateau, and overfitting is avoided by tracking the divergence of training and validation accuracy. The models outlined ahead, and the training of such models is performed on TensorFlow 1.13, using Nvidia Titan RTX Graphics Processing Units. 

\subsubsection{SegNet}
\label{sec:segnetArchitecture}

\begin{figure}
  \centering
    \includegraphics[width=\textwidth]{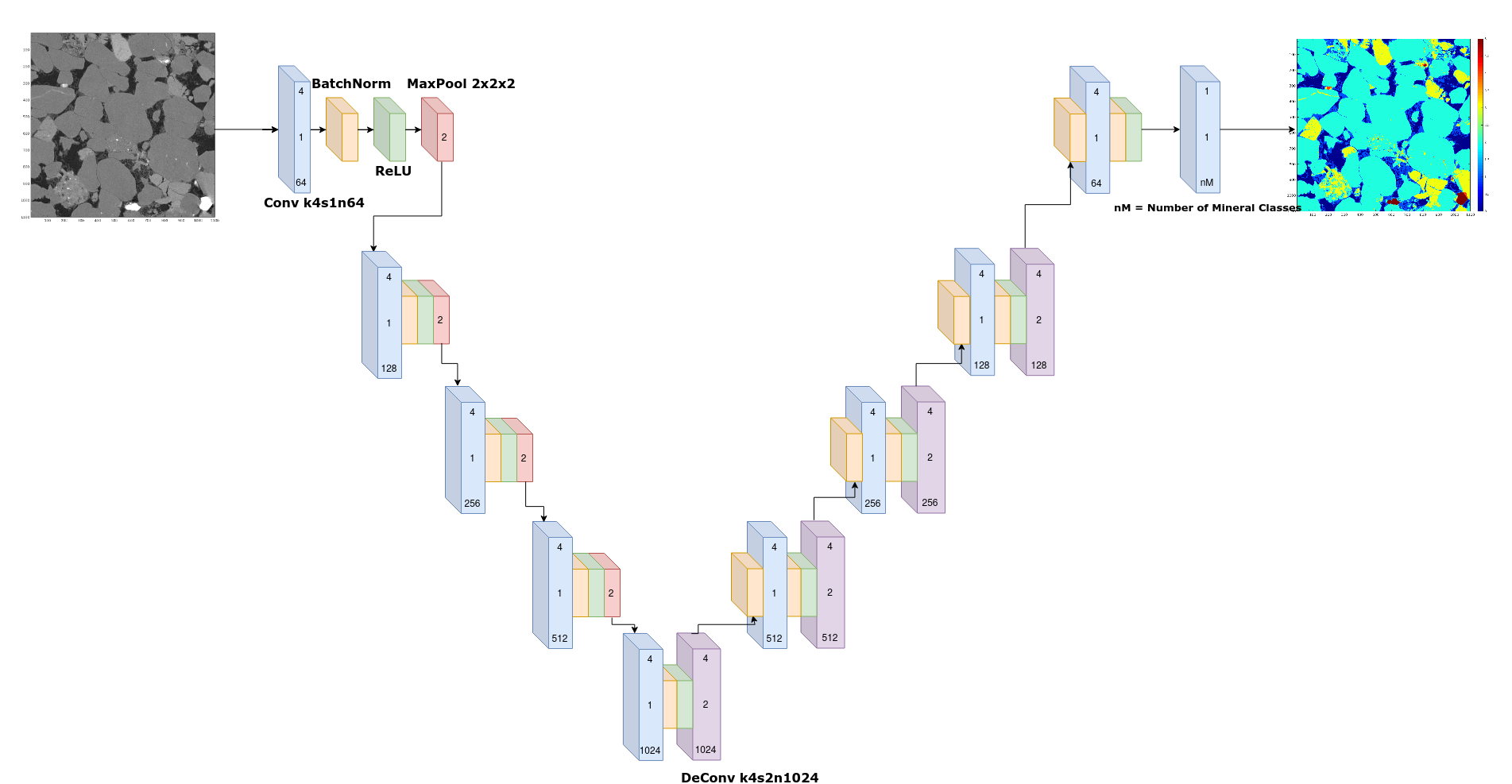}
    \caption{Architecture of SegNet, an early example of a semantic segmentation neural network. The encoding and decoding blocks are clearly illustrated. The translation of input is performed by sequentially deeper convolutions, while transformation into output is performed by sequentially shallower convolutions.}
    \label{fig:segnet2d}
\end{figure}

The SegNet architecture is comprised of an encoder-decoder structure that applies convolutional layers in sequentially downsampling steps for the encoding process, and sequential upsampling steps for the decoding process. In this process, an input grey-scale or colour image is transformed into a semantically segmented image, with multiple labelled regions identified. The architecture is illustrated in Figure \ref{fig:segnet2d}, and shows the structure of each encoder and decoder block. Each encoder block consists of batch-normalised \cite{ioffe2015batch}, ReLU activated \cite{relu} convolutions of increasing filter depth followed by a 2x2 maxpool \cite{maxpool}. These blocks sequentially transform an input tensor by subsampling the tensor along the input dimensions and storing the transformed (by convolution) information along the orthogonal dimension. This is reversed in the decoder blocks by the use of transposed convolution, that are trained to reverse the maxpooling performed during encoding while preserving identified features. 

This formulation of SegNet differs slightly to the original, which did not use transposed convolution, but instead directly reversed the pooling by the use of maxpooling indicies \cite{segnet}. This version of SegNet also uses increasingly deep convolutional filters during encoding, while the original maintained a filter size of 64. Batch normalisation layers were also not present in the original formulation, but given the relatively consistent nature of segmentation datasets, is beneficial to the learning process, so is included in this version. In this case these changes are implemented to (a) learn further upsampling features during decoding and (b) allow modular compatibility with the other networks tested. The final convolutional layer transforms the decoded data into the multi-mineral segmentation as required by the number of identified minerals within the domain. Overall, this network contains 28M trainable parameters, which is within the range of deep CNNs formulated using the encoder-decoder structure \cite{ronneberger2015unet, wangedsrgan, resnet, pix2pix2016}, typically in the 10-100M parameter range, with over 20 convolutional layers.

\subsubsection{U-Net}
\label{sec:unetArchitecture}

\begin{figure}
  \centering
    \includegraphics[width=\textwidth]{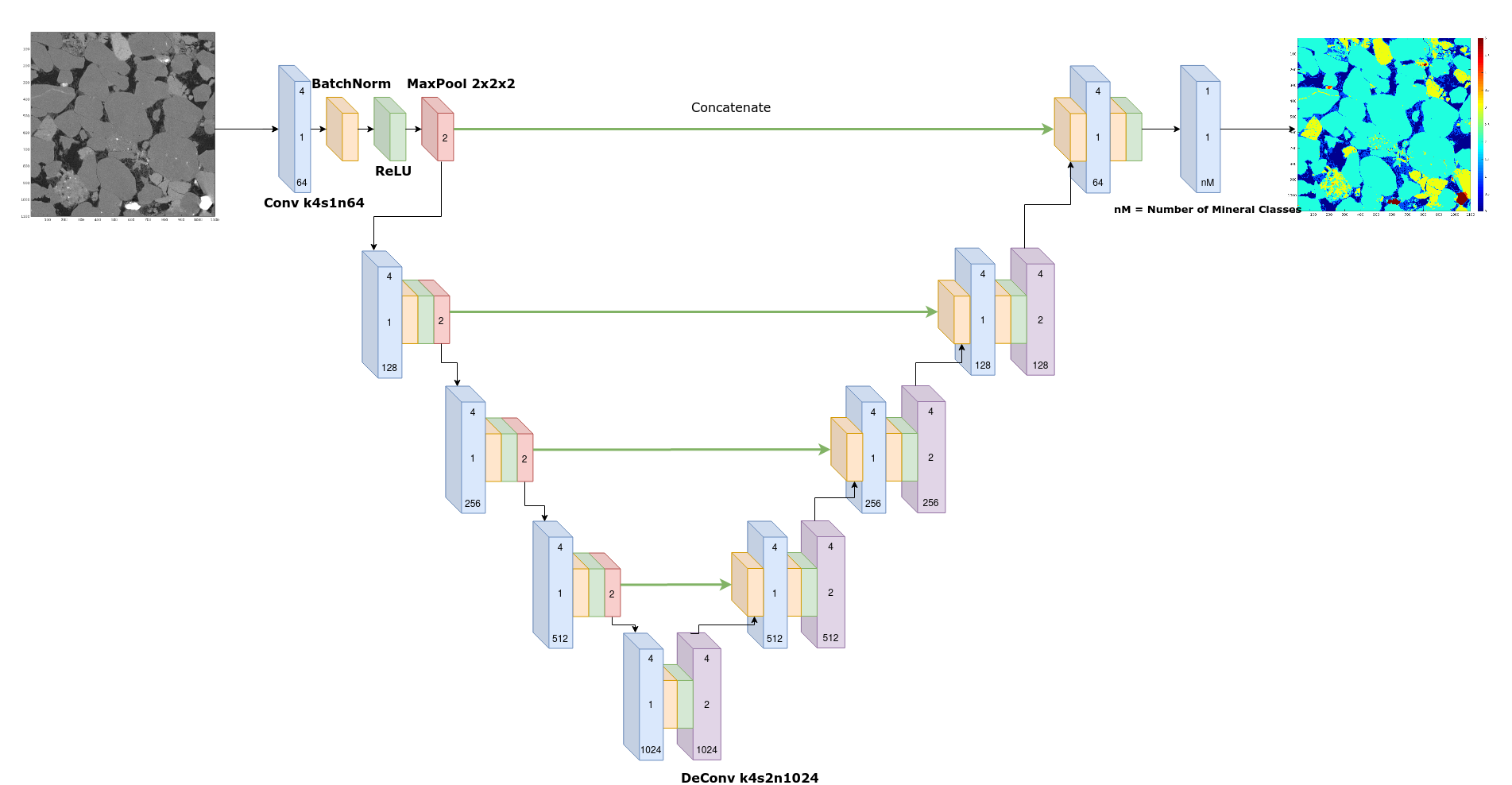}
    \caption{Architecture of U-Net, commonly used for image-to-image based segmentation oriented networks. The long skip connections used in U-net allow the preservation of small and finer scale features that would otherwise be lost during the decoding steps.}
    \label{fig:unet2d}
\end{figure}

U-net improves upon SegNet as an image-to-image translation tool \cite{pix2pix2016} primarily by the use of symmetric skip connections that link the encoder layers to their equivalent decoder layers by concatenation, as seen in Figure \ref{fig:unet2d}. This concatenation allows the shallow geometric features of the input image to be retained more readily during the decoding layers, and significantly improves detail retention during feature identification, resulting in a higer overall pixelwise accuracy. Due to the manner in which the basic SegNet is used, all of the other training parameters are preserved when using this U-net implementation.  Overall, this network contains 33M trainable parameters. 

\subsubsection{ResNet}
\label{sec:resnetArchitecture}

\begin{figure}
  \centering
    \includegraphics[width=\textwidth]{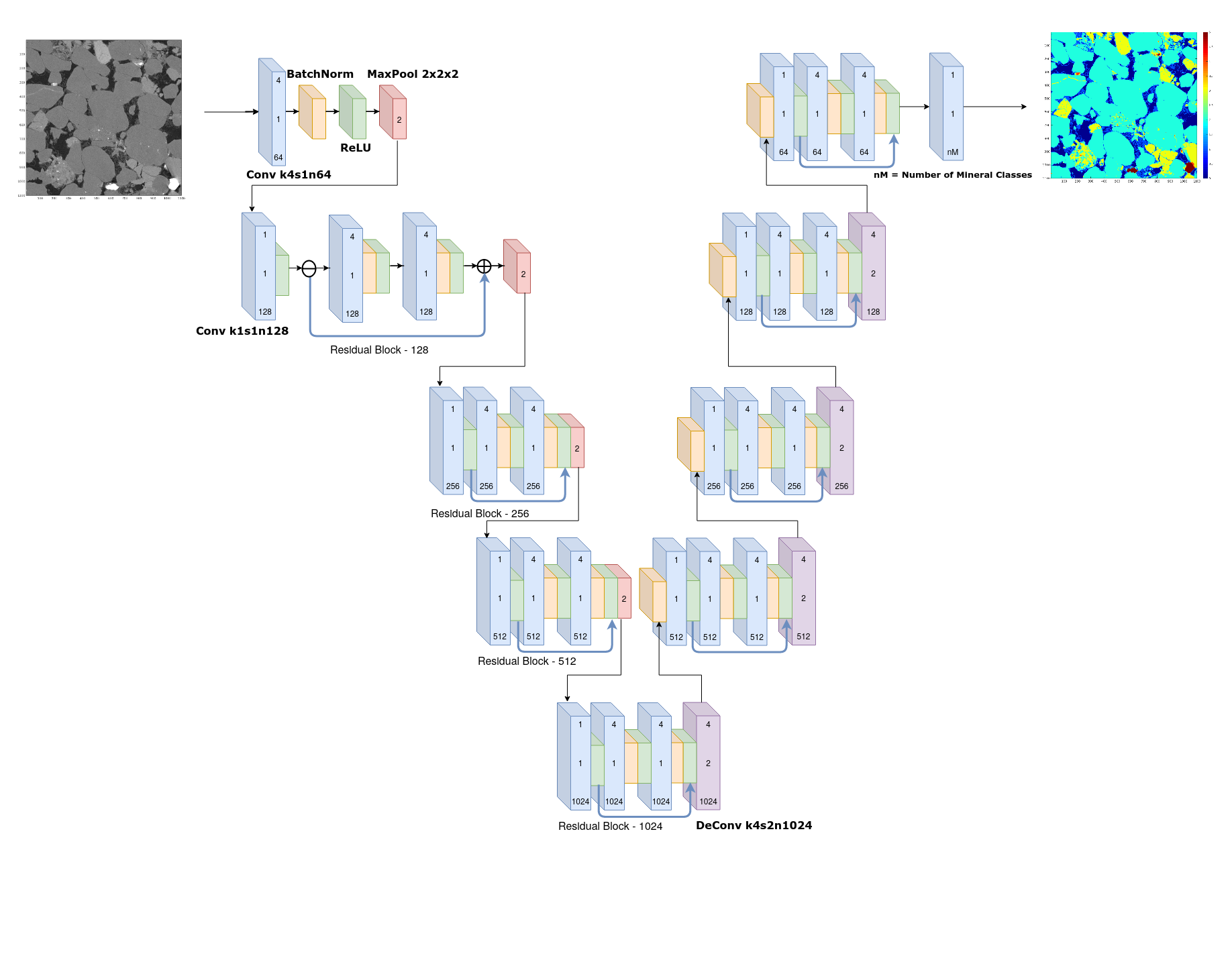}
    \caption{Architecture of ResNet as used in this study, which utilises short skip connections and 1x1 convolutions for improved deep-layer feature transfer and scaling}
    \label{fig:resnet2d}
\end{figure}

ResNet in its canonical form does not take the shape of a symmetric Encoder-Decoder, but instead is built with a much larger encoding section in the form of repeated residual blocks chained together with short skip connections. It is primarily used in image classification \cite{resnet} and in a modified form for super resolution \cite{SRGANledig}, with some use cases in image-to-image translation \cite{cyclegan}. The primary strengths of the ResNet architecture is the use of the aforementioned skip connections between blocks that allows the network to scalably and stably improve in accuracy with the depth of the network, which would otherwise not be possible \cite{resnet}. It is this short skip connection in ResNet that is appropriated to form the ResNet architecture shown in Figure \ref{fig:resnet2d}. The ResNet residual blocks of a pair of convolutions and a skip connection are also added to the architecture, and the skip connection is facilitated by a 1x1 convolution at the start of the residual block to enforce tensor size equality. Overall, this network contains 68M trainable parameters.

\subsubsection{U-ResNet}
\label{sec:uresnetArchitecture}
\begin{figure}
  \centering
    \includegraphics[width=\textwidth]{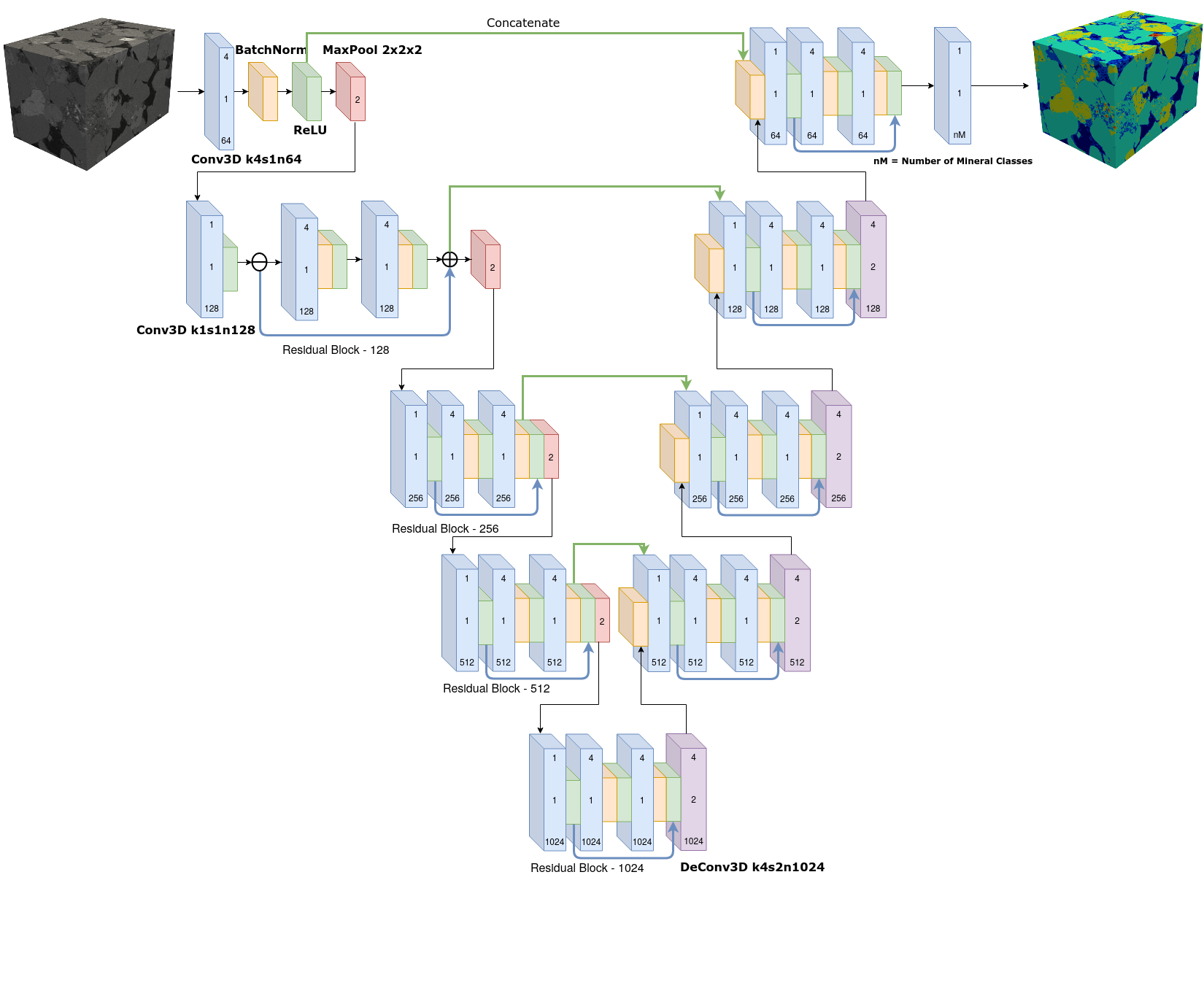}
    \caption{Architecture of U-ResNet, which combines the long skip connections in U-net with the short skip connections in ResNet. This model is also tested in 3D, hence the depiction of 3D images as input and output.}
    \label{fig:uresnet3d}
\end{figure}

The value of the skip connection is one of the primary reasons why the U-net and ResNet based architectures have outperformed other network designs. As such, a combination of the short and long skip connections that permit the networks to scale well with increasing complexity can be used to create a U-ResNet type neural network. In this case, the ResNet design in the previous section is augmented with U-net style concatenation operations between the encoder and decoder blocks. This results in the network architecture seen in Figure \ref{fig:uresnet3d}. This network is also formulated in 3D, as a 3D implementation may improve segmentation accuracy along the depth dimension, using 3D convolutions and associated operations. Overall, this network contains 68M trainable parameters in 2D, and 268M parameters in 3D.

\subsection{Loss Functions and Accuracy}
\label{sec:lossandacc}

From the output of these CNNs, the objective loss function used to train the network is the weighted softmax cross entropy, as shown in Eqn \ref{eqn:SMXE}. The softmax function transforms tensor elements $a_{i}$ in a given dimension into real numbers in range (0,1) (which add up to 1) that acts as a logit, or a "probability distribution" $p_{SM}$, that the voxel is the specified mineral, and the cross entropy $XE$ is calculated between the one-hot mineral label vectors $m$ and $p_{SM}$. For example in a segmented image measuring 100x100 pixels, with 5 identified minerals $(nM=5)$, the one-hot (sparse) segmentation map takes the form of a boolean array measuring 100x100x5. For each segmented pixel, a 5 element boolean vector represents $m$, and the resulting softmax vector $p_{SM}$ is also 5 elements long, with double precision values in the range (0,1). The cross entropy is then calculated for all $N$ vector pairs of $m$ and $p_{SM}$ over the image, in this case for a 100x100 image, $N=10,000$. 

One aspect of accuracy in using semantic segmentation for multi-mineral segmentation is the relative contributions of accuracy from each identified mineral. In a given sample, the occurrence of minerals is varied, and can range from under 1\% to over 90\%. The segmentation accuracy compared to the ground truth for each mineral should be considered, and is affected by the overall accuracy achievable by the network, dataset representation, and loss function weighting. In the case of a weighted softmax cross entropy, a vector of weights $w$ is also multiplied into the cross entropy in order to counterbalance the occurrence rate of certain mineral classes. For example, if mica minerals occurred only 2\% of the time while clay minerals occurred 20\% of the time, then the machine learning would be skewed to favour clays, as it would contribute more to the lowering of the loss function. In this case, the inverse of the bulk fraction of each mineral class is used to weigh the loss function.

\begin{equation}
\label{eqn:SMXE}
\begin{aligned}
{p}_{SM_i}=\frac{e^{a_i}}{\sum_{k=1}^{nM}e^{a_k}} \\
XE=-\frac{1}{N}\sum_{i=1}^Nwmlog(p_{SM})
\end{aligned}
\end{equation}

In order to measure the pixelwise accuracy of the segmentation, the Mean Logical Error is calculated for each phase as well as the bulk volume. This is calculated as the ratio between the total number of correctly segmented pixels and the total number of pixels of the phase of interest. Any errors in segmentation of a phase will result in further errors in the segmentation of other phases since the segmentation labels are densely populated.

\subsection{Physical Measurements of Accuracy}
\label{sec:segFlowAccuracy}

Aside from voxelwise accuracy, segmentation accuracy should be measured also by physical parameters since much of the time in Digital Rock Analysis, the objective of segmentation is to isolate a region for further physics based analysis. The topology and connectivity of the pore space and the mineral phases are important physical measures, while single phase and multi-phase flow are critical phenomena that are simulated with segmented images in the pore space of the domain. As such, they are good measures of the physical accuracy of segmentation, assuming there is some ground truth reference to compare against. Usually this is experimentally obtained values such as permeability, relative permeability, or insitu phase distribution. In this case, the ground truth is assumed to be the manual segmentation by converging-active-contours that the dataset is derived from. In a generalised use case, where the availability of manually segmented multi-mineral images is abundant, this would be a reasonably unbiased and accurate representation of the "ground truth".

\subsubsection{Connectivity}
\label{sec:Connectivity}
The connectivity and morphology of phases as identified by segmentation is an important physical parameter that affects subsequent flow analysis in the pore space while grain contact distribution affects the solid mechanics of the digital image. Connectivity and morphology can be measured by the Euler Number $\chi$, calculated as the difference between the number of disconnected bodies ${\chi}_{0}$ and the number of loops ${\chi}_{1}$ such that ${\chi} = {\chi}_{0}-{\chi}_{1}$ \cite{Armstrong2019}. 

\subsubsection{Single-Phase and Multi-Phase Flow}
\label{sec:singlemultiphase}

Flow within the pore space is calculated by the Lattice Boltzmann Method (LBM) using a Multi-Relaxation Time (MRT) scheme in D3Q19 quadrature space \cite{wang2019multi}. LBM reformulates the Navier-Stokes Equations (NVE) by numerically estimating the resulting continuum mechanics from underlying kinetic theory. The kinetics of a bulk collection of particles within a control volume is estimated with a 19 vector velocity space $\xi_{q}$ and velocity distributions $f_{q}$. For each velocity space vector $\xi_{q}$, the velocity component in the specified direction is given by $f_{q}$. Using these concepts, an equation can be constructed that details the development of fluid transport. In particular, the momentum transport equation at location $\vec{x}_{i}$ over a timestep $\delta t$ takes the form in Eqn \ref{eqn:lbmMomentum} that relies on a collision operation $J$ which is model specific and outlined in detail in \cite{MCCLURE20141865}:

\begin{equation}
\label{eqn:lbmMomentum}
f_{q}(\vec{x}_{i}+\vec{\xi_{q}}\delta t, t+\delta t) = f_{q}(\vec{x}_{i}, t) + J(\vec{x}_{i}, t)
\end{equation}

Single phase flow is simulated within the pore space of the segmented test samples to obtain the sample permeability. This is obtained by simulation to steady state conditions and calculating the permeability $K$ by Eqn \ref{eqn:perm}:

\begin{equation}
\label{eqn:perm}
K=\frac{\mu \bar{\vec{v}}L}{\nabla P_{x}}
\end{equation}

where $\mu$ is the kinematic viscosity, $\bar{\vec{v}}$ is the mean velocity within the bulk domain, $L$ is the length of the sample in the direction of flow, and $\Delta P$ is the pressure difference between the inlet and outlet.



Multi-phase flow is solved within the pore space using Colour Gradient LBM, of which the specific implementation is detailed in \cite{MCCLURE20141865, wang2019multi}, and is performed as primary drainage injection of Non-Wetting Phase (NWP) into the sample that is initially fully saturated with Wetting Phase (WP). This process is continued until the NWP reaches the outlet side of the domain. The system is assumed to be uniformly wetting with a contact angle of 45 degrees, driven by a viscous force that results in a Capillary Number $N_{C_A}$ of 5E-5. It should be noted that, since the domain is segmented per mineral, the system wettability can be modelled as a mixed-wetting system. A uniform wetting is assumed to simplify the simulation and analysis process. A mixed wetting system is modelled in Section \ref{sec:mixedCompare} to illustrate the complete workflow for multi-mineral segmented images.

During this simulation, the topology of the NWP is tracked as it is forcibly injected into the pore space. The physical accuracy compared to the ground truth is measured as the deviation between both the topology of the pore space, and the topology of the NWP phase, as both of these parameters differ. 


\section{Results and Discussion}
\label{sec:Results and Discussion}

In total, the networks outlined in Section \ref{sec:segNets} in 8 different configurations are tested in 2D, and the best performing is then tested in 3D. Furthermore, the influence of dataset representation is acknowledged, so the training is performed in weighted and unweighted formats as described in Eqn \ref{eqn:SMXE}, where $w_{i}=1$ for unweighted and $w_{i}=\frac{1}{\phi_{i}}$ for weighted where $\phi_{i}$ is the \% volume fraction for each mineral $i$. Thus 10 different results are obtained from 5 different neural networks, of which 4 are in 2D and 1 is in 3D. The results obtained on the testing section of the dataset are analysed both voxelwise and based on their physical attributes.

\subsection{Multi-Mineral Segmentation Accuracy}
\label{sec:accuracy2D}
\begin{figure}
  \centering
    \includegraphics[width=\textwidth]{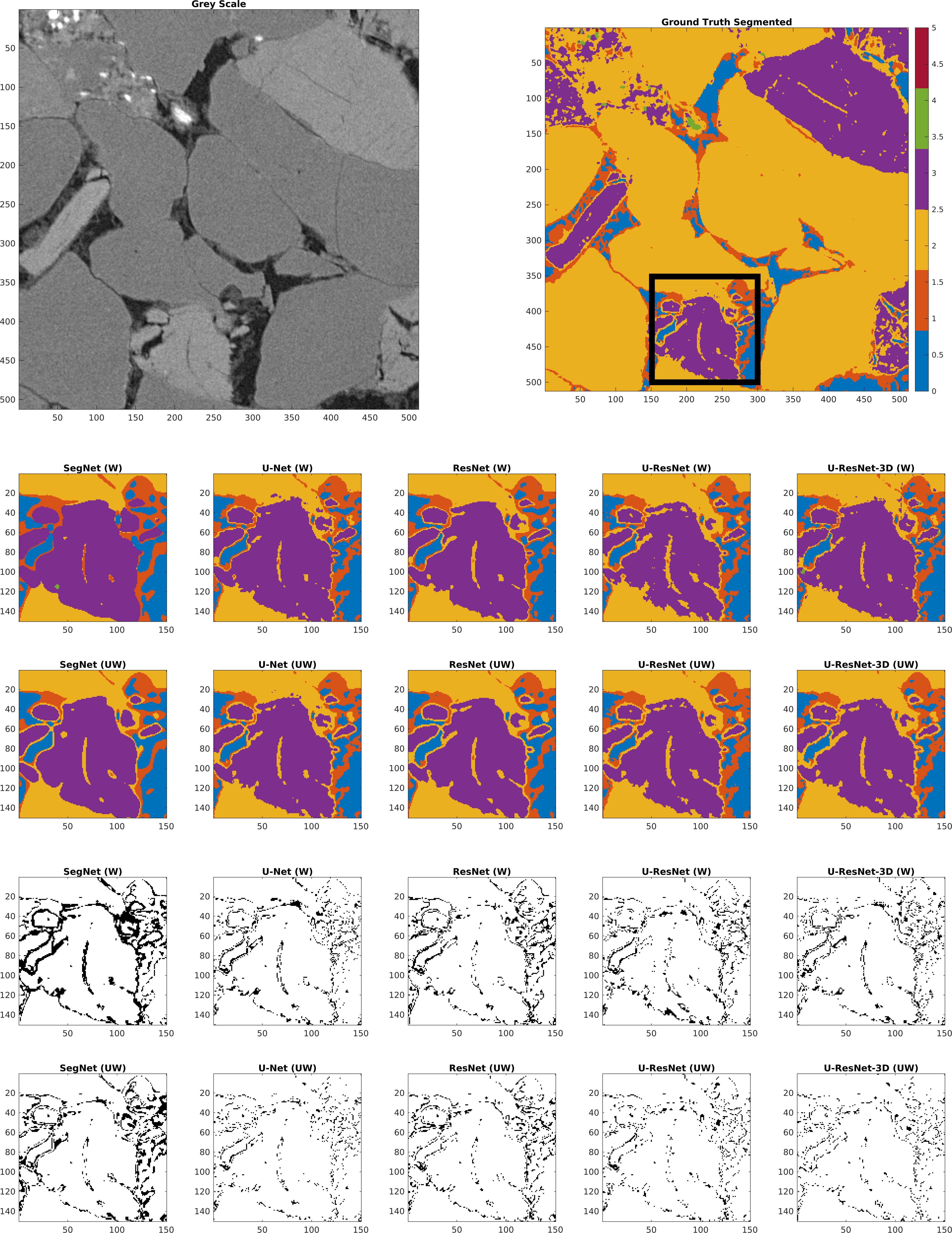}
    \caption{Top: A slice from the testing dataset. Middle: Sample visualisations of the segmentation results as obtained from the network configurations tested with and without weighting (W = weighted and UW = unweighted), compared to the region of interest identified by the black bounding box in the top right. Bottom: Corresponding regions of segmentation error showing that the unweighted U-ResNet and U-Net results visually perform best, consistent with Table \ref{tab:segNetsAccuracyTable}}
    \label{fig:segNetFullSamples}
\end{figure}

The pixelwise accuracy of the segmentation is calculated on a testing sample measuring 512x512x768 voxels, obtained from the testing dataset. In all tests shown in this study, this volume is used, which represents roughly 10\% the entire dataset. In 2D, segmentation is generated slice-by-slice, while in 3D, blocks of 128x128x128 are stitched together to form the volume. This stitching in 2D and 3D is to ensure that, while the sample being used is a single volumetric domain, it is comprised of a wide selection of subsamples as generated by the trained Neural Networks. This unseen section of the dataset is segmented with the 10 models tested and detailed in Table \ref{tab:segNetsAccuracyTable}, with visualisation of a region of interest shown with accuracy maps in Figure \ref{fig:segNetFullSamples}. Overall, the U-ResNet architecture performed best in 2D, and 3D testing further outperforms in most cases, and in terms of pore space segmentation, which is the most critical phase for conducting flow simulation, the best results were generated by the U-ResNet-3D unweighted network. The U-net architecture comes in a close second, as expected given its design and popular use as the architecture of choice on most applications \cite{ronneberger2015unet, pix2pix2016, cyclegan}. These accuracy results are supported visually by Figure \ref{fig:segNetFullSamples}, showing in the error maps that the least error occurs within the U-ResNet and U-Net segmentation results. 

\begin{table}
\centering
 \caption{Accuracy results for identified phases in the 512x512x768 testing sample using different segmentation trained neural networks with and without weighting (W = weighted and UW = unweighted). The best and second best accuracy results are highlighted in \textcolor{red}{red} and \textcolor{orange}{orange} respectively. Overall, the U-ResNet architecture performed best in 2D, and 3D testing further outperforms in terms of voxelwise accuracy.}
 \label{tab:segNetsAccuracyTable}
  \begin{tabular}{|c|c|c|}
    \hline
     & Mean Acc. & Weighted Mean Acc.\\
    \hline
    \hline
    SegNet (W) & 	0.9075 &	0.8693 \\
    \hline
    UNet (W) & 	0.8540 &	0.9436 \\
    \hline
    ResNet (W) & 	0.8814 &	0.9188  \\
    \hline
    UResNet (W) & \textcolor{red}{0.9450} &	0.9282 \\
    \hline
    SegNet (UW) & 	0.8563 &	0.9166 \\
    \hline
    UNet (UW)  & 	0.8929 &	0.9556  \\
    \hline
    ResNet (UW) & 	0.8765 &	0.9385  \\
    \hline
    UResNet (UW) & 	0.8738 &	\textcolor{orange}{0.9598}\\
    \hline
    \hline

    UResNet-3D (UW) & \textcolor{orange}{0.9365} &	0.9348 \\
    \hline
    UResNet-3D (W) & 0.8908 &	\textcolor{red}{0.9677} \\
    \hline
    \hline
  \end{tabular}
    \begin{tabular}{|c|c|c|c|c|c|c|}
    \hline
     &  Pore &  Clay & Quartz & Feldspar & Micas & Mixed  \\
    \hline
    Vol (\%) & 0.0693 & 0.1136 & 0.6103 & 0.2046 & 0.0021 & 0.0002 \\
    \hline
    SegNet (W) & 0.9564 &	0.8369 &	0.8325 &	0.9677 &	0.8644 &	\textcolor{red}{0.9874} \\
    \hline
    UNet (W)  & 0.9573 &	0.9239 &	0.9639 &	0.8912 &	0.7622 &	0.6252 \\
    \hline
    ResNet (W) & \textcolor{orange}{0.9876} &	0.8497 &	0.9116 &	0.9556 &	\textcolor{orange}{0.9334} &	0.6506 \\
    \hline
    UResNet (W)  & 0.9791 &	\textcolor{orange}{0.9482} &	0.8978 &	\textcolor{red}{0.9909} &	0.9090 &	\textcolor{orange}{0.9448} 	\\
    \hline
    SegNet (UW) & 0.9554 &	0.7694 &	0.9329 &	0.9385 &	0.7466 &	0.7950 \\
    \hline
    UNet (UW)   & 0.8783 &	\textcolor{red}{0.9552} &	\textcolor{orange}{0.9682} &	0.9456 &	0.8911 &	0.7194 \\
    \hline
    ResNet (UW) & 0.9651 &	0.8424 &	0.9535 &	0.9409 &	0.6909 &	0.8664 \\
    \hline
    UResNet (UW) & 0.9810 &	0.9154 &	0.9601 &	0.9774 &	0.9009 &	0.5080 \\
    \hline
    \hline

    UResNet-3D (UW) & \textcolor{red}{0.9936} &	0.8770 &	0.9209 &	\textcolor{orange}{0.9882} &	\textcolor{red}{0.9538} &	0.8857 	\\
    \hline
    UResNet-3D (W)  & 0.9750 &	0.9204 &	\textcolor{red}{0.9738} &	0.9753 &	0.7995 &	0.7007 	\\
    \hline
    \hline
  \end{tabular}
\end{table}

The volume fraction of each mineral phase is important to note, since a high accuracy achieved in phases of lesser importance and occurrence can come at the cost of accuracy in more commonly occurring and more important phases. The high accuracy achieved in the segmentation of sparsely occurring minerals (by volume fraction) using SegNets is primarily due the SegNet producing segmentation realisations that lack edgewise detail. The sparsely occurring minerals tend to exist in small bodies only a few voxels in radius that the SegNet captures by overestimating the segmented body as a large blob. This in turn reduces the segmentation accuracy of other phases. This effect of sparse label accuracy bias is due to both the limitations of the network as well as the occurrence rate of examples within the training dataset (volume fraction). 

By using loss function weighting based on the inverse of the volume fraction, the weighted results (denoted as (W)) show a more even accuracy spread over the phases, achieving higher accuracy in segmentation of sparsely occurring minerals. This however, in turn reduces the segmentation accuracy of other more commonly occurring phases. The choice of weighted or unweighted training should be based on both the importance of certain phases as well as the volume fraction within the dataset.


In general, of the networks tested, the U-net and the U-ResNet architectures achieved the best results likely due to their use of skip connections to preserve fine detail. The variation in accuracy between the networks designed with skip connections (ResNet, U-Net, and U-ResNet) all perform several percentage points higher than SegNet, which has mostly been superseded. The 3D U-ResNet results are marginally superior to the 2D results, which again is expected for a dataset generated by 3D segmentation, where the relationship between phases in the depth dimension is important. For even the sparsest occurring phases, U-ResNet and U-net achieve high accuracy in the 80-90\% range without significantly sacrificing accuracy achieved in segmentation of other phases.

\subsection{Physical Accuracy as Measured by Connectivity}
\label{sec:eulerPhases}
\begin{figure}
  \centering
    \includegraphics[width=\textwidth]{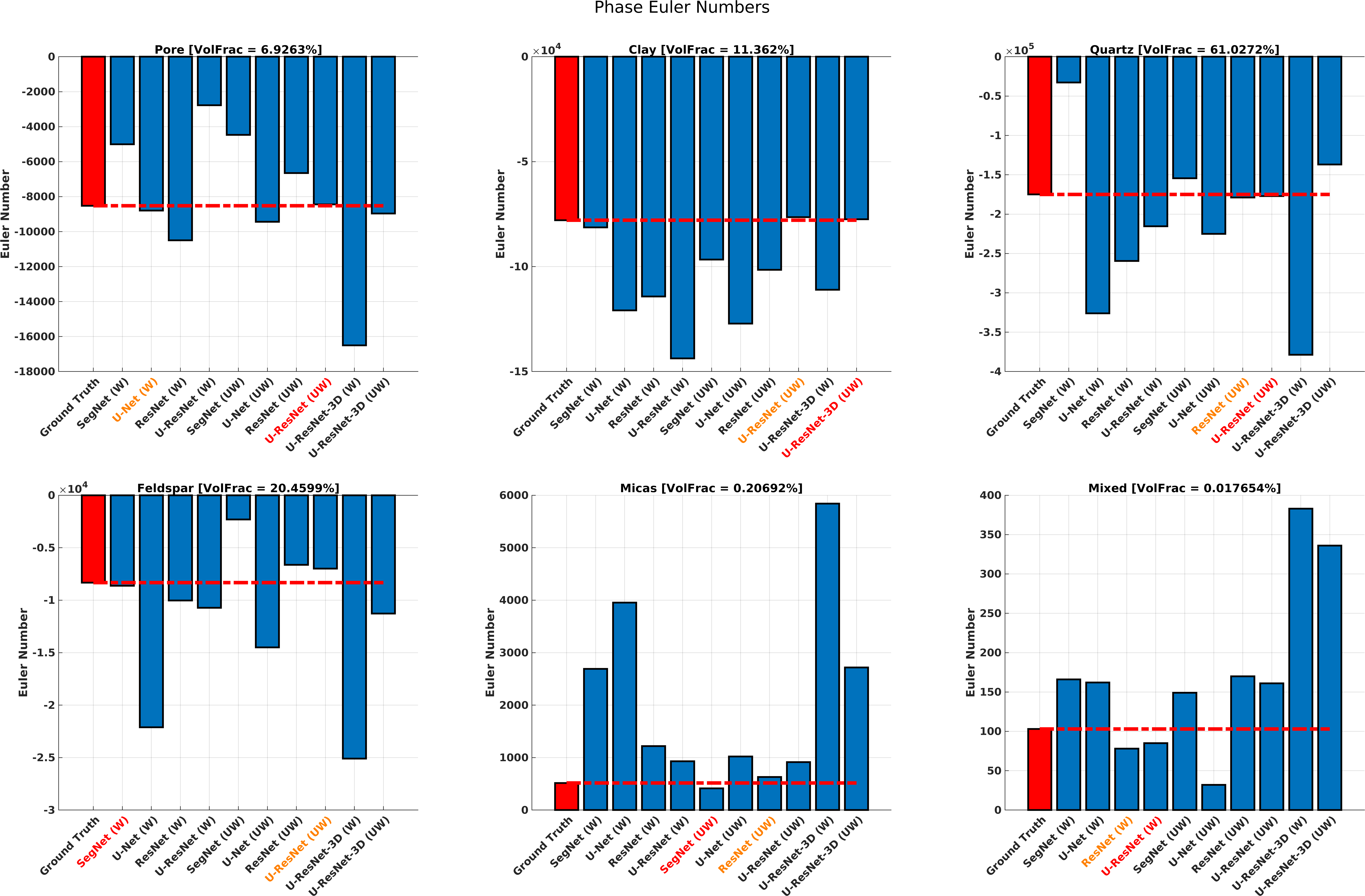}
    \caption{Barchart plots of the calculated Euler Number on each segmented phase for each of the 10 tested networks, compared to the ground truth result. The best and second best accuracy results are highlighted in \textcolor{red}{red} and \textcolor{orange}{orange} respectively. The connectivity is erratic compared to previously calculated pixelwise accuracies and no single network is able to reliably generate the same connectivity.}
    \label{fig:chiSegNets}
\end{figure}

Aside from the pixelwise accuracy (vs Ground Truth segmentation) achieved by 2D and 3D segmentation neural networks the physical accuracy of the segmented realisations are arguably a more critical metric to consider. In the context of Multi-Mineral segmentation of digital rock images, one of the main physical parameters to consider is the topology/connectivity of each phase. As an extension of the pixelwise measures of segmentation accuracy, the Euler number of each segmented phase is also calculated as described in Section \ref{sec:Connectivity} on the 512x512x768 testing sample. The results are shown in Figure \ref{fig:chiSegNets} as bar charts over the 6 phases of the 11 samples that comprise the Ground Truth sample, as well as the 10 samples generated by the trained networks.

The connectivity achieved on each phase by each realisation varies significantly, with errors reaching 1000\% in some cases. This erratic variation in the Euler Number implies that the connectivity of the segmented phases is not reliably preserved during the training and testing phase. The presence of convolutional artefacts and/or single-pixel errors located near the boundaries of loops and disconnected bodies are the likely reasons behind the large variability in physical connectivity. Scattered artefacts tend to increase the Euler number, while pixel errors near the boundaries of larger bodies can form loops that decrease the Euler number. Regardless, looking at the best and second best performing networks, as well as the general trends of network accuracy measured by Euler number, the U-ResNet variants tested tend to perform most reliably. In terms of the connectivity of the pore space, U-ResNet (UW) and U-net (W) performed best, followed by U-ResNet-3D (UW). This is in contrast to the pixelwise segmentation accuracy measures, where U-ResNet-3D (UW) performed best. 

While some networks perform better in general when measured by Euler Number accuracy, no single tested network is entirely successful in preserving the segmented connectivity of all phases. In the case of the 3D U-ResNet tests, U-ResNet-3D (UW) performs adequately within range and magnitude for all predominant phases, with the sparsely occurring micas and mixed mineral phases showing high relative error as trade-off for the lower error in more abundantly occurring phases.

\subsection{Comparison of Computed Permeability and Multi-Phase Topology}
\label{sec:permCompare}
\begin{figure}
  \caption{}
  \label{fig:permSegNets}
  \centering
    \includegraphics[width=\textwidth]{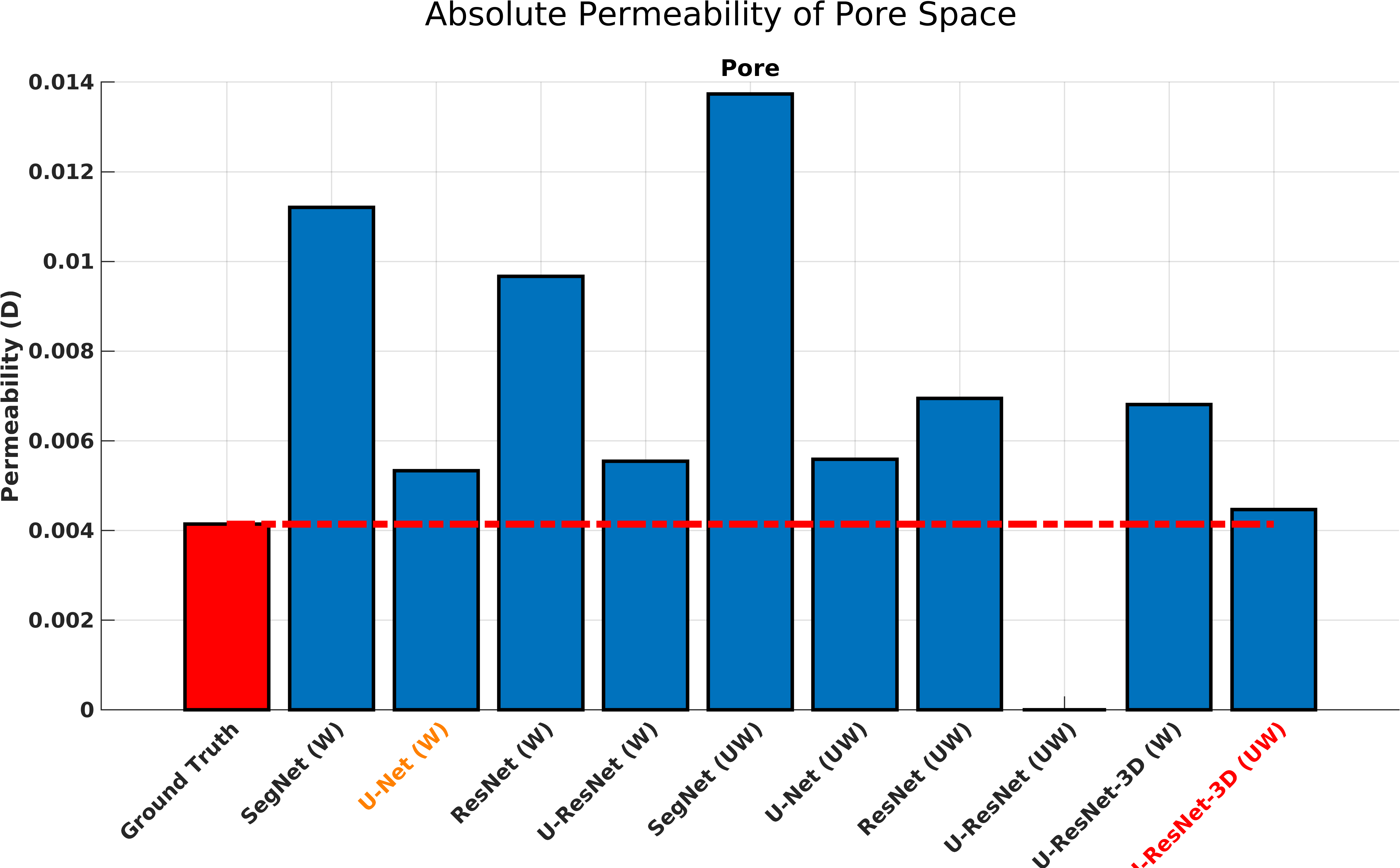}
  \hfill
    \caption{Bar chart comparison of the absolute permeability, with closest match achieved by the U-ResNet-3D (UW). All other results overestimate significantly, illustrating  higher sensitivity to physical measures of accuracy}
\end{figure}

The flow characteristics of the pore space within the segmented images is a critical measure of physical accuracy, with both single and two fluid phases. In the case of single phase flow, the absolute permeability of the pore space is the primary parameter to determine, and is done so as outlined in Section \ref{sec:singlemultiphase} using LBM-MRT. The same 512x512x768 testing sample as used in previous sections is used to simulate within the pore space until steady state conditions are reached. The calculated permeabilities are reported in Figure \ref{fig:permSegNets} and show by bar chart comparison of the simulated absolute permeability by LBM-MRT at steady state conditions. 

\begin{figure}
  \centering
    \includegraphics[width=\textwidth]{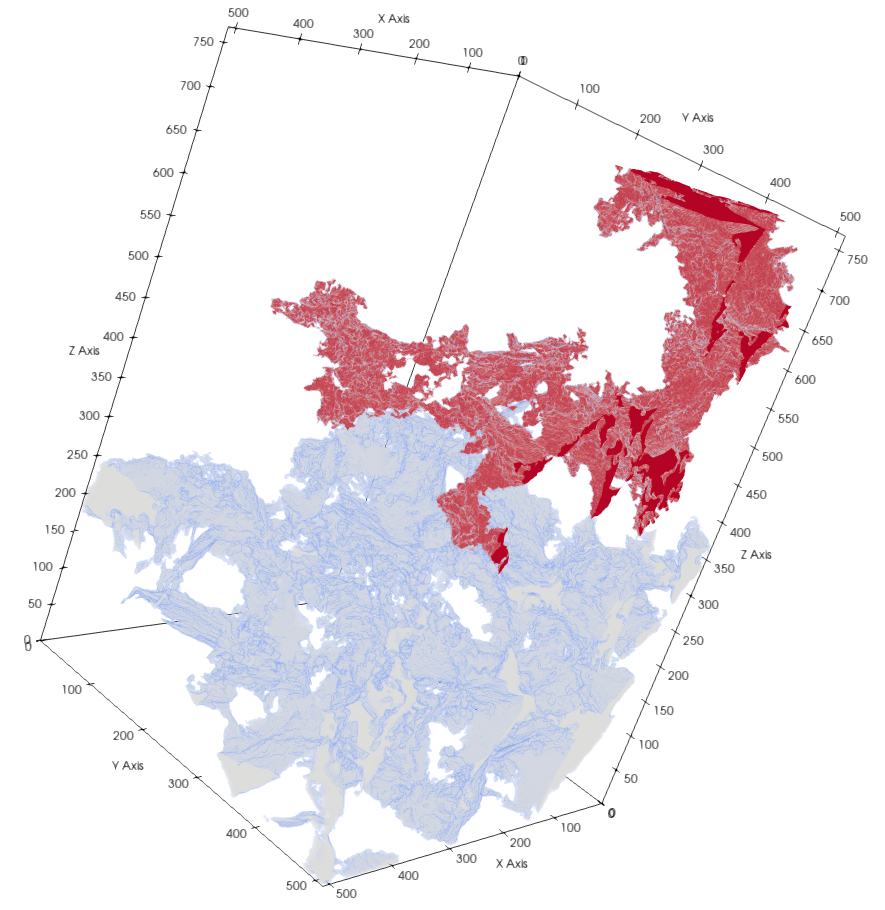}
    \caption{Visualisation of the expected flow path in the testing sample. This disconnected pore space is highlighted by the 2 disconnected bodies in blue and red, which is connected in the original Ground Truth, but disconnected in the U-ResNet-2D result that corresponds to a zero permeability shown in Figure \ref{fig:permSegNets}}
    \label{fig:disconSimon}
\end{figure}

\begin{figure}
  \centering
    \includegraphics[width=\textwidth]{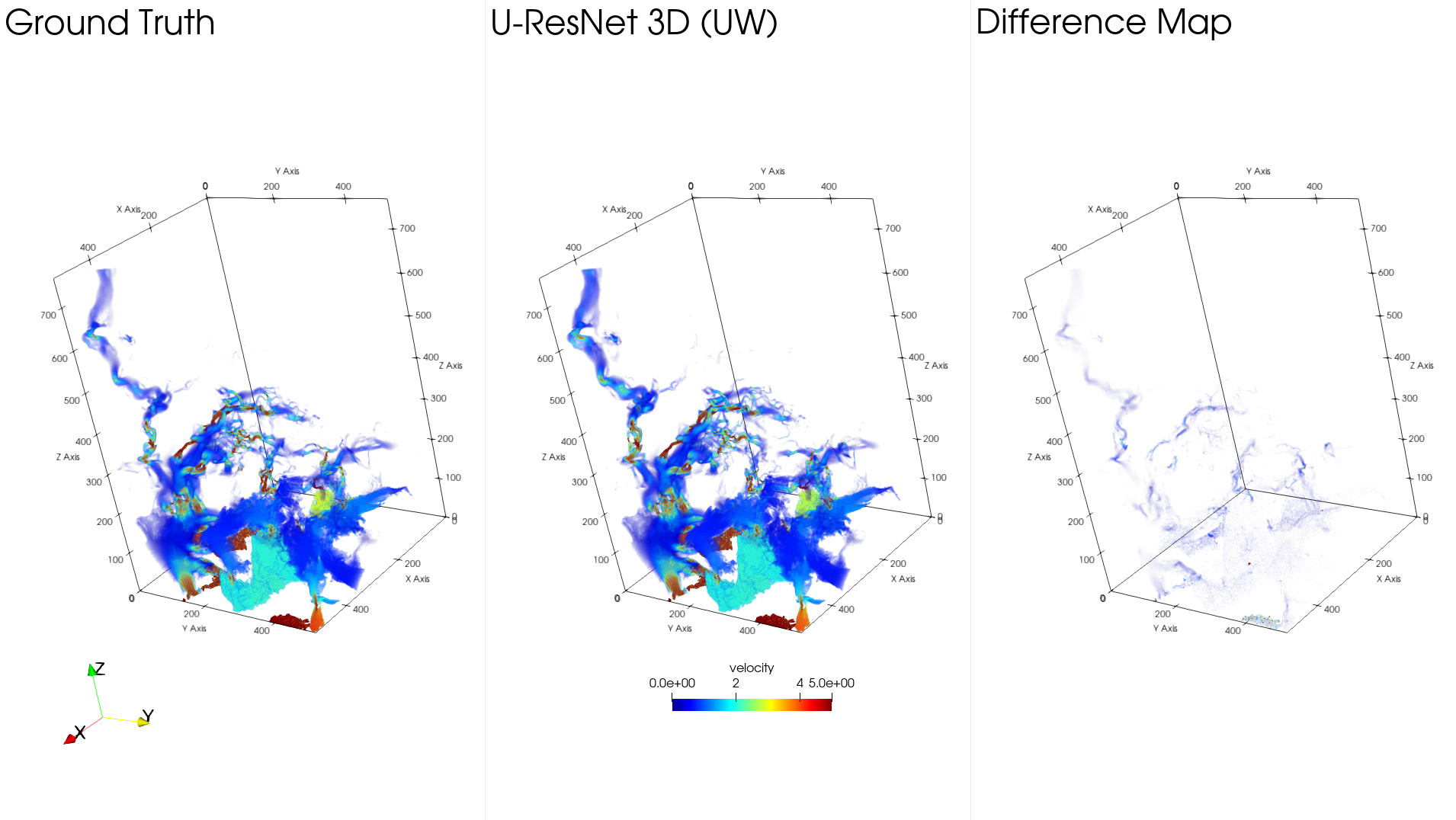}
    \caption{Comparative visualisation of velocity magnitude field as obtained from direct simulation using LBM on the pore space on a 512x512x768 voxel sample from the 3D testing dataset. The resulting images are visually similar, but the discrepancy in calculated permeability can be seen in various locations where the U-ResNet result simulates a higher local velocity. A clear example can be seen when comparing the top flow path}
    \label{fig:segCell111VelMag}
\end{figure}

The closest match is achieved by the U-ResNet-3D (UW) result, which is supported by this also having the highest pixelwise accuracy in the segmentation of the pore space. All other results are off by an order of magnitude, illustrating the higher sensitivity to physical measures of accuracy. Of interest is also the zero permeability result obtained on the U-ResNet (UW) result, indicating that the pore space is disconnected by the segmentation applied. This is also despite having the best Euler Number match in the pore space as seen in Figure \ref{fig:chiSegNets}, which indicates that the Euler number is not entirely sufficient as a physical measure of accuracy. This particular testing sample is quite heterogeneous, and the flow path is particularly tight in the outlet area of the sample, as can be seen in Figure \ref{fig:segCell111VelMag}. The location of this disconnection can be seen in Figure \ref{fig:disconSimon}, and appears to occur at a tight pore throat junction. This again reinforces the need to verify segmentation performance by neural networks with physical measures alongside pixelwise measures. The best match achieved by the U-ResNet-3D (UW) result is visualised in Figure \ref{fig:segCell111VelMag}, showing a good overall match in simulated velocity fields, with minor differences due to variations in the pore space. 

\begin{figure}
  \centering
    \includegraphics[width=\textwidth]{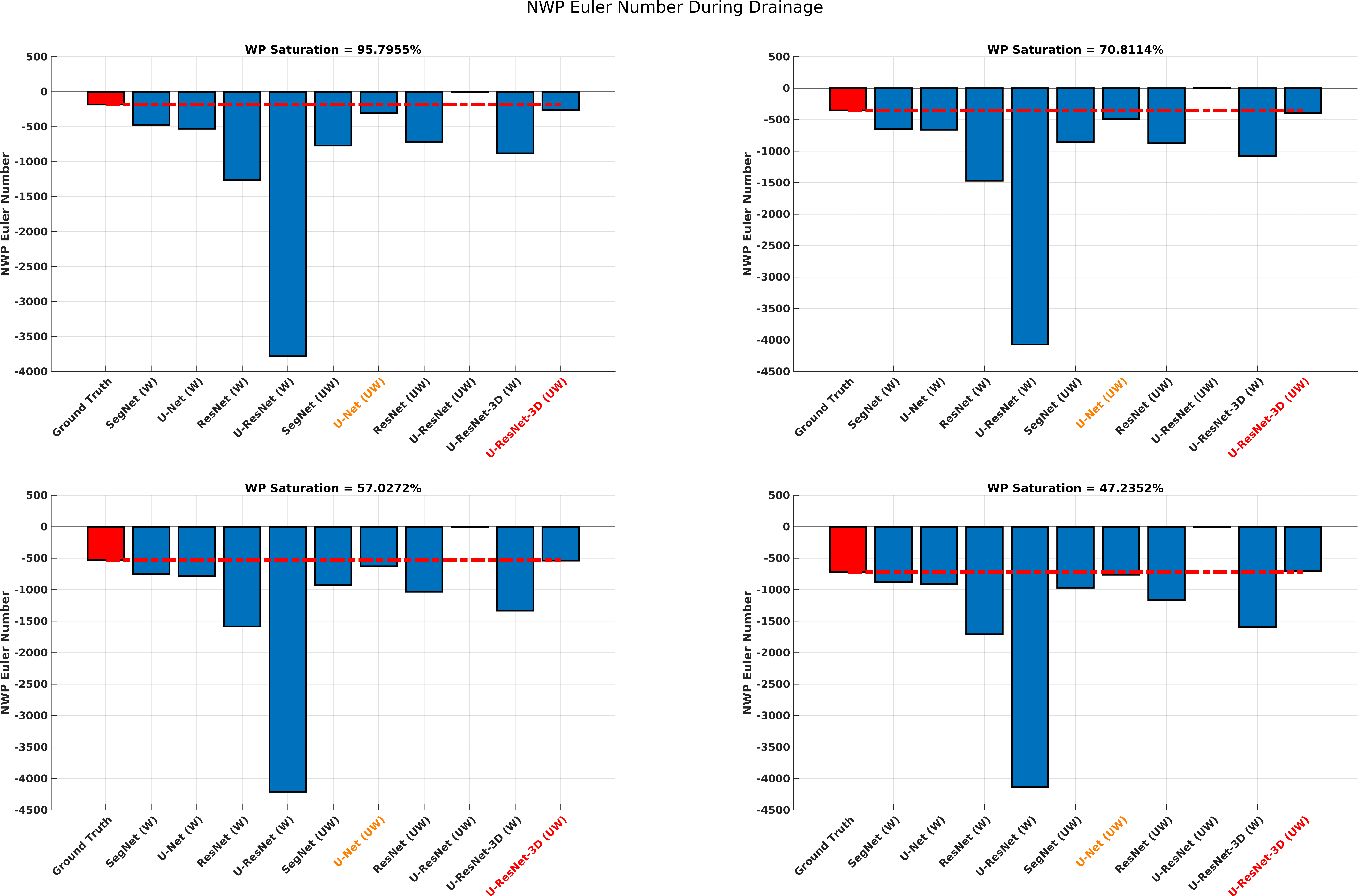}
    \caption{Barcharts detailing the evolution of the NWP Euler Number during primary drainage as simulated by multiphase LBM. As is consistent with previous tests of accuracy and physical accuracy, the U-ResNet-3D (UW) and U-Net results most closely match.}
    \label{fig:nwpChiSegNets}
\end{figure}

\begin{figure}
  \centering
    \includegraphics[width=\textwidth]{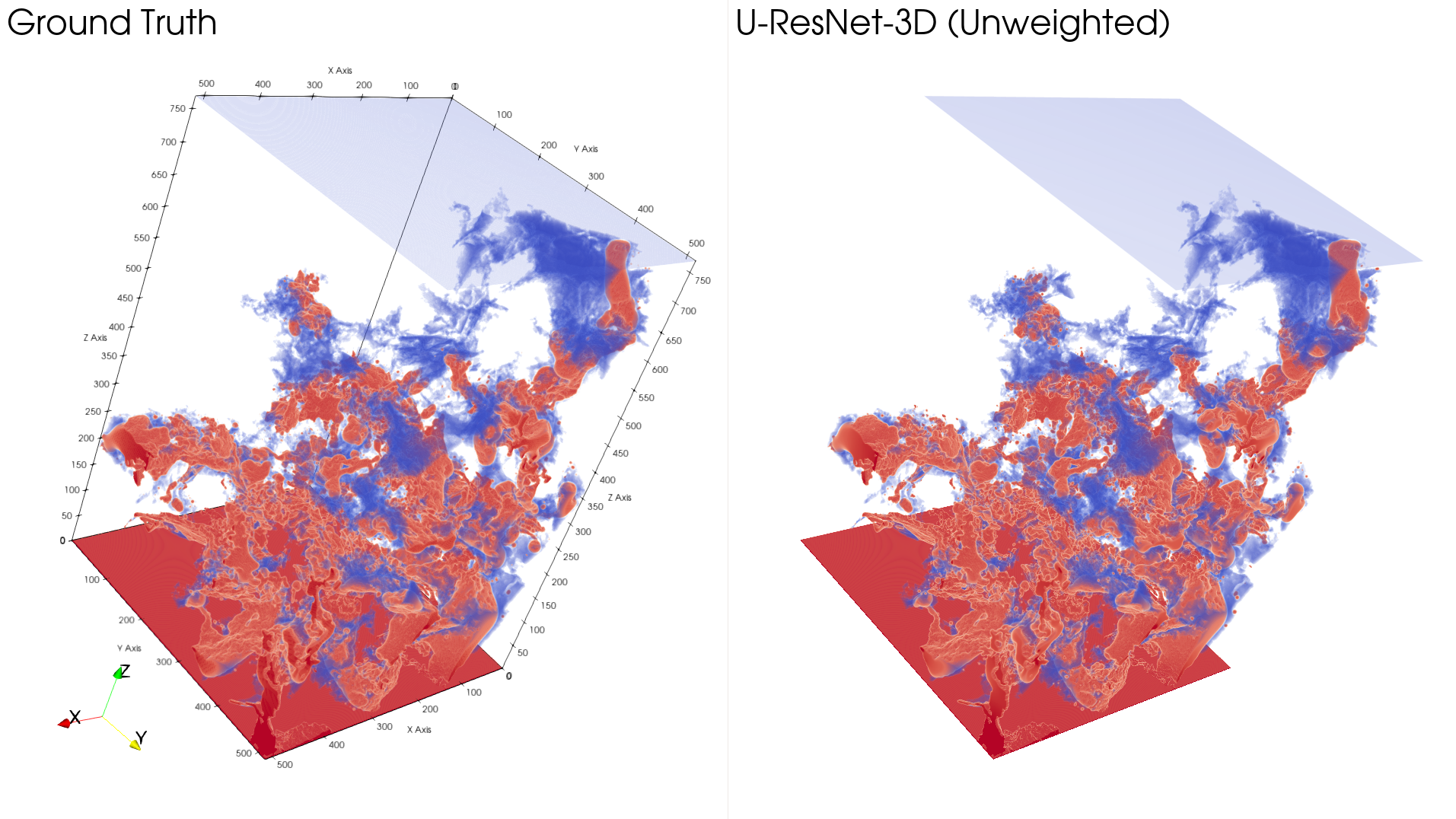}
    \caption{Visualisation of the phase distribution after 400,000 LBM timesteps simulating primary drainage, reaching breakthrough at 47\% saturation. When comparing the original result on the ground truth geometry and the best performing U-ResNet-3D (UW) results, a close match can be observed. Blue corresponds to WP, and red to NWP}
    \label{fig:segCell111Drain45400000}
\end{figure}

Multiphase flow is equally if not more important in many cases, since the pore space can contain oil-water or water-air mixtures in varying proportions. The simulation of multiple phases within the pore space of resolved porous media (such as digital rocks) is challenging both in terms of formulation and computation, but is absolutely necessary as a tool in the digital analysis of petrophysics. In the case of the connectivity of NWP flowing within the pore space, the geometry of the pore space closely affects the resulting configuration of phase within \cite{snapoff}. A drainage injection test is simulated within the pore space of the 11 segmented images, all under the same conditions, and the evolution of the phase distribution is tracked by calculating the Euler Number of the NWP. This number represents the connectivity of the injected phase over the course of the simulation, which runs from being fully saturated with WP, to first breakthrough of NWP at the outlet. The NWP Euler Number is plotted for the 11 samples tested, at LBM timestep intervals of 100,000 as the NWP invades the pore space, in Figure \ref{fig:nwpChiSegNets}. The results show clearly that the U-ResNet-3D (UW) and U-Net results most closely match, consistent with previous tests of accuracy and physical accuracy. The disconnected result from the permeability tests in Figure \ref{fig:permSegNets} is also disconnected here, and much like the case in connectivity of mineral phases in Section \ref{sec:eulerPhases}, convolutional artefacts and pore boundary errors are likely the primary reasons why certain networks perform better or worse. The best simulation results as obtained on the geometry generated by the trained U-ResNet-3D (UW) is visualised in direct comparison with the ground truth simulation in Figure \ref{fig:segCell111Drain45400000}, and a close match can be observed. 

\subsection{Multi Phase Topology Comparison under Mixed Wetting Conditions}
\label{sec:mixedCompare}
\begin{figure}
  \centering
    \includegraphics[width=\textwidth]{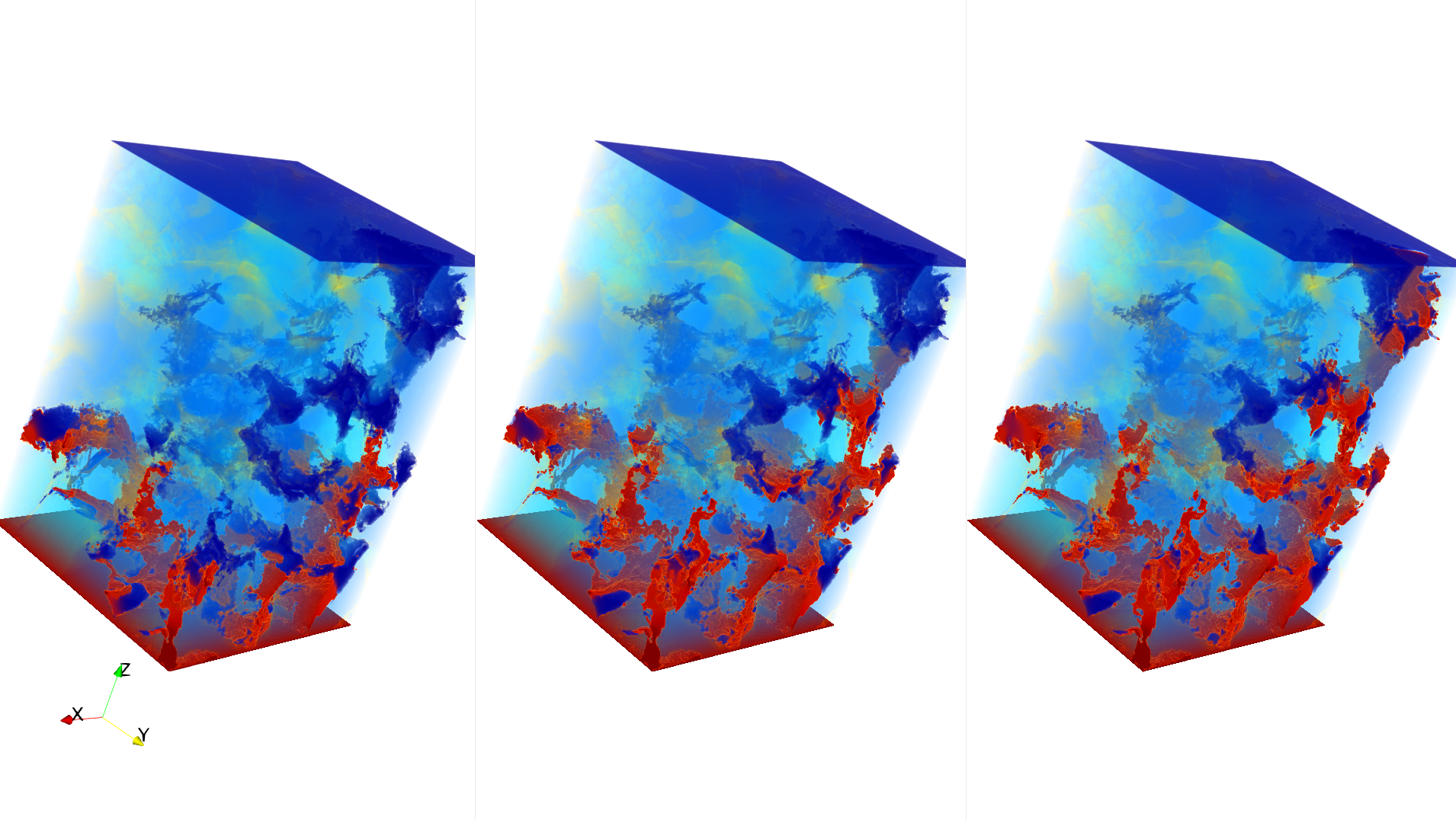}
    \caption{Visualisation of primary drainage in a mixed wetting system as simulated by multiphase LBM. Blue corresponds to WP, and red to NWP. Each phase with a different contact angle is shown with opacity, with light blue = Feldspar, dark blue = Quartz, and yellow = clay. Compared to the simulation of a single-wetting system, the phase distribution of NWP (w.r.t quartz) is more disordered, transitioning from pore bodies to pore walls depending on the local wettability, resulting in a higher number of loops (seen in the NWP Euler number in Figure \ref{fig:nwpMixedChiSegNets}) forming due to imbibition and drainage phenomena occurring to the same fluid body at the same time.}
    \label{fig:nwpMixedVis}
\end{figure}

\begin{figure}
  \centering
    \includegraphics[width=\textwidth]{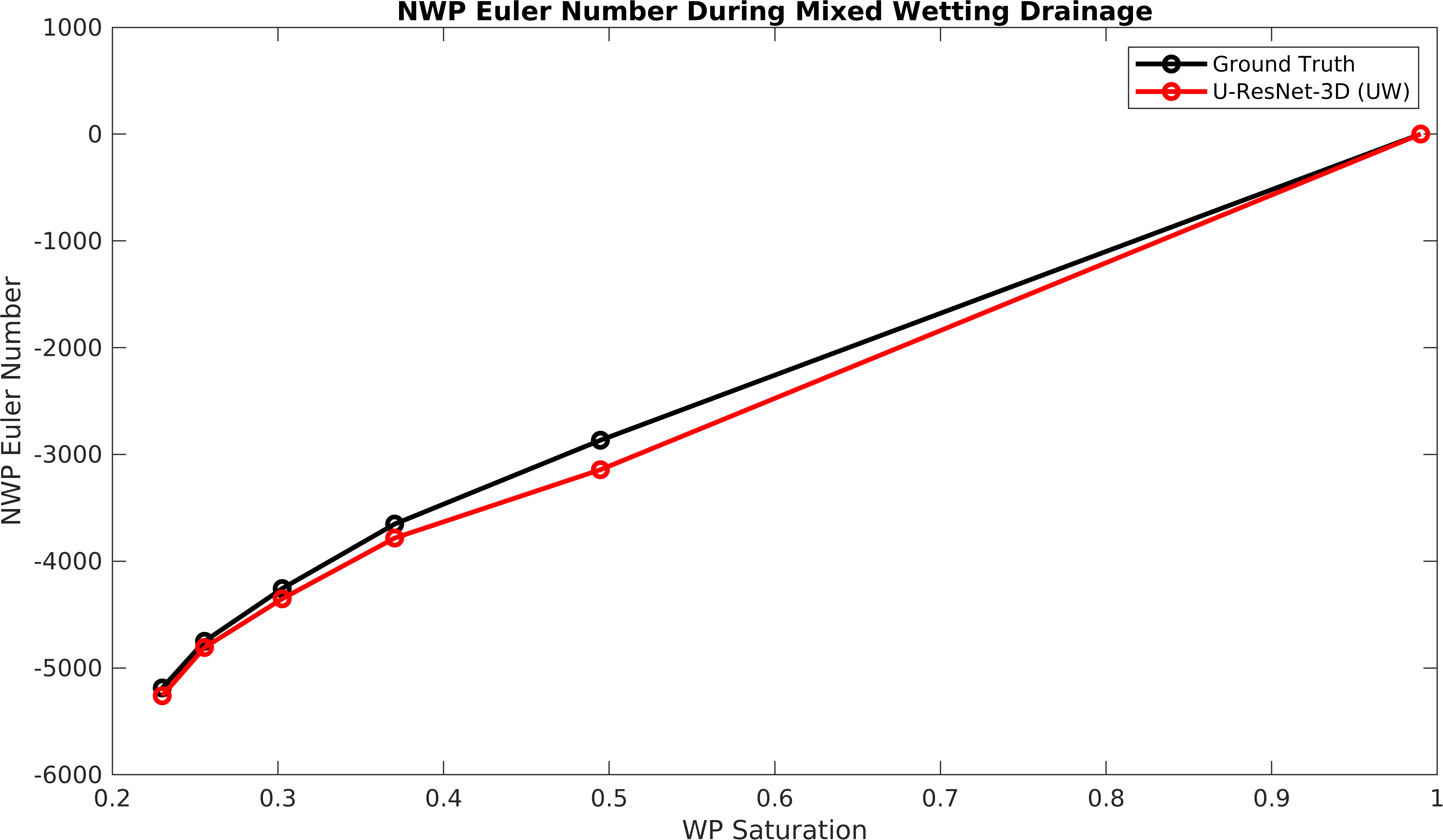}
    \caption{The evolution of the NWP Euler Number during primary drainage in a mixed wetting system as simulated by multiphase LBM. This mixed wetting simulation test is performed on the best performing result, as generated by U-ResNet-3D (UW). As is consistent with previous tests of accuracy and physical accuracy, the U-ResNet-3D (UW) closely matches the ground truth.}
    \label{fig:nwpMixedChiSegNets}
\end{figure}

Having established that the 3D U-ResNet Unweighted network performs the most consistently in terms of both pixelwise and physical accuracy measures, a comparison is now made on the segmentation accuracy as measured by direct simulation of mixed-wetting multiphase flow within the pore space. This test represents the workflow that completely preserves the resulting segmentation, and makes no simplifications to the petrophysical analysis of multi-mineral segmented samples. The ground truth and U-ResNet-3D Unweighted segmentations are used as input domains for a primary drainage simulation in a mixed wetting environment. Using multiphase LBM, the input contact angles are assigned based on typical values of the minerals identified. The clays are assigned a contact angle of 110 degrees, the quartz 60 degrees, and the feldspar 70 degrees. The other mineral labels are sparsely populated, so are assigned a neutral contact angle of 90 degrees. All other simulation parameters are identical to the single wetting tests performed in the previous section. The NWP (with respect to quartz) is tracked and analysed topologically by the Euler number during the simulation. A rendering of the simulation is shown in Figure \ref{fig:nwpMixedVis}, and the Euler number evolution throughout the simulation is shown in Figure \ref{fig:nwpMixedChiSegNets}. Overall, due to imbibition and drainage phenomena occurring to the same fluid body at the same time, the phase distribution of NWP (w.r.t quartz) is more disordered, transitioning from pore bodies to pore walls depending on the local wettability, resulting in a higher number of loops forming.

It can be seen that, as is consistent with previous tests of physical accuracy, the resulting segmented geometry from the U-ResNet-3D Unweighted network conforms consistently to the ground truth in physical accuracy, in this case deviating by between than 2\% to 9\% in Euler number throughout the simulation, which implies that the resulting segmentation of mineral phases (and subsequent contact angles) closely matches the ground truth.

\section{Conclusions}

Multi-Mineral segmentation (and segmentation in general) is critical in the Digital Rock Analysis workflow, and CNN based segmentation can reduce unwanted bias and manual tuning required in conventional segmentation methods. The SegNet, U-Net, ResNet, and U-ResNet CNN models tested in 10 configurations all show relatively high pixelwise accuracy, though the U-Net and U-ResNet architectures perform best in this regard. The importance of maintaining physical accuracy is emphasised, by applying further downstream analysis to measure physical accuracy. These physical measures of connectivity, permeability, and phase distribution show significantly higher variance (over 1000\% in some cases) with models that otherwise achieves 95\%+ in voxelwise accuracy. The newly introduced U-ResNet-3D network architecture as a hybrid fusion of U-net and ResNet outperforms all other tested models in voxelwise and physical accuracy measures, with the traditional U-Net following closely behind. When using CNN based segmentation methods, the network architecture should be carefully considered, and validation should extend to testing of physical accuracy to ensure usability.

This particular study does not encompass all manner of CNN architectures, and does not cover other methods that are highly automated and designed to reduce bias. The models used and proposed are designed to illustrate the improvement in accuracy (both pixelwise and physical) that may be obtained with increasing CNN complexity. It can be expected that, with deeper and different architectures, a potentially even higher accuracy can be achieved. Furthermore, certain measures of physical accuracy such as connectivity or an approximation of flow may be calculable or estimated in an efficient manner that is conducive to use directly as a loss function to train neural networks to specifically minimise the error in physical accuracy. The mass availability of data is critical and key for utilising these CNN methods accurately, as the use case shown in this study is confined to a localised dataset segmented manually by human biases. It would take a larger dataset of many different varieties of rock samples all multi-mineral segmented by different individuals to truly remove subjectivity, and the methods and findings in this study would assist in guiding the training and testing of such a model. 

\section{Acknowledgements}
The source code used in this study available at \url{https://github.com/yingDaWang-UNSW/SegNets-3D}.

\bibliographystyle{plain}  

\end{document}